\documentclass[a4paper,10pt]{article}
\usepackage[pdftex]{graphicx,hyperref}
\usepackage{amsmath,amsfonts,amssymb}
\usepackage{wasysym}
\usepackage{fullpage}
\usepackage{enumerate}
\usepackage{sidecap}
\usepackage{lscape}
\usepackage{authblk}
\usepackage{appendix}
\usepackage{setspace}
\usepackage{algorithm}
\usepackage{clrscode3e}
\usepackage{algpseudocode}
\usepackage{pseudocode}
\usepackage{subcaption}
\title{\bf{User-based representation of time-resolved multimodal public transportation networks}}
\author[1,2,3]{Laura Alessandretti}
\author[1]{M\'arton Karsai \thanks{marton.karsai@ens-lyon.fr}}
\author[2]{Laetitia Gauvin}
\affil[1]{\small{Laboratoire de l'Informatique du Parall\'elisme, INRIA-UMR 5668, IXXI, ENS de Lyon, 69364 Lyon,France}}
\affil[2]{\small{Data Science Lab, ISI Foundation, Torino, Italy}}
\affil[3]{\small{Department of Mathematics, City University London, London EC1V0HB, United Kingdom}}

\begin{document}

\maketitle

\begin{abstract}
Multimodal transportation systems can be represented as time-resolved multilayer networks where different transportation modes connecting the same set of nodes are associated to distinct network layers. Their quantitative description became possible recently due to openly accessible datasets describing the geolocalised transportation dynamics of large urban areas. Advancements call for novel analytics, which combines earlier established methods and exploits the inherent complexity of the data. Here, our aim is to provide a novel user-based methodological framework to represent public transportation systems considering the total travel time, its variability across the schedule, and taking into account the number of transfers necessary. Using this framework we analyse public transportation systems in several French municipal areas. We incorporate travel routes and times over multiple transportation modes to identify efficient transportation connections and non-trivial connectivity patterns. The proposed method enables us to quantify the network's overall efficiency as compared to the specific demand and to the car alternative.
\end{abstract}

\section{Introduction}

Urban transportation systems interweave our everyday life and although their construction is based on conscious design they appear with complex structural and dynamical features \cite{citeulike:13661976}.  They build up from different transportation means, which connect places in a geographical space. Their most straightforward description is given by networks \cite{sheffi1985urban, citeulike:6861223} where stations are identified as nodes and links are the transportation connection between them. Based on this representation \cite{PhysRevE.67.036106} considerable research efforts have been dedicated to address their sustainability \cite{citeulike:250373} and to optimise their efficiency \cite{Mandl1980396, Banavar1999Size}, and reliability \cite{citeulike:13662280,citeulike:13662281, Oort2009Regularity}. These investigations have been accelerated lately by the availability of large open datasets describing complete multimodal transportation systems in cities, regions, countries, and even internationally. These advancements became possible due to the novel techniques of data collections, and amplified by the actual trends in urban policy propagating smart city programs and real time online user services.

All transportation networks share a few common features: (a) they are all embedded in space, setting constraints in their structural design, (b)
networks of different transportation means may coexist in the same space, and (c) they are all inherently temporally-resolved. The recent developments in data collection practices and in the corresponding fields of complex networks and human dynamics provided the opportunity to study these systems quantitively using a data-driven approach. These studies showed that geographical constraints largely determine the structure and scaling of transportation networks \cite{doi:10.3141/2112-03,PhysRevE.72.046127,10.1371/journal.pone.0029721} but for their better understanding one needs to consider the actual urban environment and development level \cite{Banavar1999Size,10.1371/journal.pone.0102007,citeulike:13662014}. At the same time the emerging field of multilayer networks provided the methodology to consider their multimodal character \cite{citeulike:13266661, citeulike:13590899}. In this representation each layer corresponds to the network of a single transportation mean (bus, tram, train, etc.), which are defined on the same set of nodes (stations). This way they account for possible multiple links of different modes between the same stations \cite{citeulike:13661002}. This representation can be extended to capture the temporal nature of the system by using some aggregated information extracted from the transportation schedule \cite{citeulike:13480203} or, as a future challenge, by considering each time slot as a layer where journeys between stations are represented as temporal links \cite{citeulike:13266661, citeulike:10864311}.

In this paper we build on these contemporary advancements to analyse large transportation networks of French municipalities represented as time-resolved multilayer networks. Our aim is to provide a novel user-based framework to represent public transportation systems by incorporating the minimisation of the total travel time, its variability across the schedule, and taking into account the number of transfers between lines. Such framework allows to identify non-trivial connection patterns in the transportation network and to quantify the network's overall efficiency as compared to the specific demand and to the car alternative. The datasets we use are openly shared by local public transportation companies operating in the larger Paris area, Toulouse, Nantes, and Strasbourg. The collected data corpora are in General Transit Feed Specification (GTFS) format and contains the geographically detailed informations about transportation trips and routes, together with travelling times, frequencies and transfer times for each service line and station.

As follows first we describe the actual time-resolved multilayer network representation and introduce a methodology incorporating travel routes and times to identify efficient transportation connections as compared to travels by cars. Next we apply matrix factorisation methods to extract underlying connectivity patterns to analyse them from the commuter point of view, and quantify their overall efficiency. Finally we conclude our results and discuss possible applications and future directions of research. Note that the implementation of the proposed methodology is openly accessible online\footnote{\url{https://github.com/lalessan/user_basedPT}}.

\section{Representation of public transportation networks}

The proposed methodology integrates several sequential steps to detect origin-destination areas that are conveniently connected by public transportation with respect to user preferences. In the following description, first we define a user-based representation of a Public Transportation (PT) system, which limits the effect of its spatial embeddedness, but accounts for its multilayer structure, and its temporal dimension. Next, we calculate shortest time paths between stops by adapting a conventional algorithm \cite{citeulike:180287} to the actual graph representation, and finally we select preferred connections, taking into account distance traveled over time. 

\subsection{User-based multi-edge P-space representation}

Several studies revealed that the choice of users to select transportation means for commuting is mainly affected by the average travel time, and by the variability of the total travel time \cite{asperges2007determinants, le2010mobilite}, in addition to the number of transfers they need to do. Other less determinant factors, that we do not consider here, include travel cost and comfort. Our representation of the PT network incorporates the aforementioned aspects decisive for users and minimises the effects due to the spatial embeddedness of the system. In order to do so we combine a multi-edge \cite{thesis:MultiColored} and a P-space representation of the transportation network \cite{paper:PspaceIndian,paper:PspacePoland,paper:PspaceComplete} to describe the PT systems. The multi-edge representation accounts for the presence of several transportation lines in the same PT network by allowing the existence of multiple labelled edges within a single pair of nodes.
On the other hand, the P-space representation takes into account that transfers between lines is time-consuming and may not be convenient for the user; also, it considers connections between stops located at large distance thus it reduces the effect of the geographical distances. The combination of these two representations for modelling public transportation systems constitutes an ideal framework to investigate their complex features from the user perspective. A schematic example of this representation is displayed in Fig. \ref{SchemaPaths}.

\begin{figure*}[htbp!]
\centering
\includegraphics[width=\textwidth]{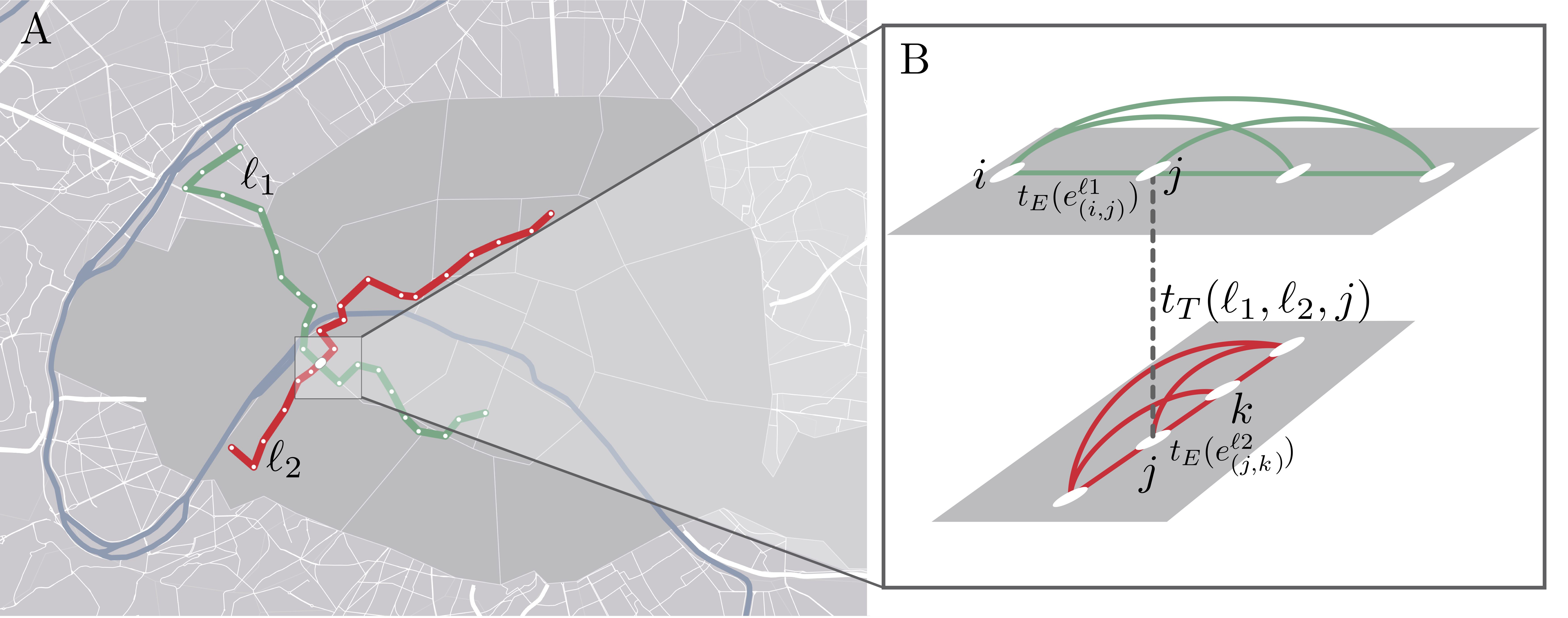}
\caption{\textbf{Illustration of the user-based multi-edge P-space representation.} A) Two geo-localised crossing PT lines $\ell_1$ and $\ell_2$ are shown on the map of central Paris. B) Schematic illustration of the P-space multi-edge representation for a section of the network: all pairs of nodes corresponding to stops on the same line are connected by edges with the same label. For example, nodes $i$ (resp. $k$) and $j$ on the same line $\ell_1$ (resp. $\ell_2$), are linked through the edge $e^{\ell_1}_{(i,j)}$ (resp $e^{\ell_2}_{(j,k)}$), with weight $t_E(e^{\ell_1}_{(i,j)})$ (resp. $t_E(e^{\ell_2}_{(j,k)})$). At the interchange node $j$, there exists a transfer with weight $t_T(\ell_1,\ell_2,j)$ allowing to change line.}
\label{SchemaPaths}
\end{figure*}

Formally, the public transportation system is defined as a weighted, directed, edge labelled graph $G=(V, E, t_E, T, t_T)$ with vertex set
$V$ with cardinality $N$, corresponding to the public transportation stops, edge set $E$ with weight function $t_E$, and set of transfers $T$ with weight function
$t_T$. If $\ell_k$ is defined as an ordered sequence of stops connected consecutively by a single transportation line, in the corresponding
P-space graph $G$ there will be a direct labelled-edge $e_{ij}^{\ell_k} \in E$  connecting any pair of nodes $(i,j)$ such that stop $i$ proceeds 
stop $j$ in the sequence of line $\ell_k$. We define $M$ as the total number of lines in the PT system. Further, a 
set $T \subset M \times M \times N$ of transfers identifies triplets of two lines and one node, $e^T_{\ell_1,\ell_2,j}=(\ell_1,\ell_2, j)$ such 
that a transfer between lines $\ell_{1}$ and $\ell_{2}$ is possible at station $j$. Each edge in $E$ is weighted by the average travel time on
the line considered, computed as
follows. This average and those following are made on a selected time window of the schedule. We introduce a $time$ function $t_{E}: E \rightarrow \mathbb{R}^{+} $, quantifying for each edge $e_{ij}^{\ell_{k}}$ the time needed to 
get from $i$ to $j$ along the line $\ell_k$ averaged on the selected time window. The travel time linked to 
an edge $e_{ij}^{\ell_k} \in E$ is then calculated as the sum of the average waiting time and the average time spent on the vehicle 
as $t_{E}(e_{ij}^{\ell_k})=\frac{1}{2f_{\ell_{k}}}+ \Delta t_{ij}^{\ell_k}$ where $f_{\ell_k}$ is the average frequency of line $\ell_k$ and
$\Delta t_{ij}^{\ell_k}$ is the average time one needs to spend on line $\ell_k$ to go from stop $i$ to stop $j$.
This formula is designed to consider the case where a user would go blindly to a stop (without looking at the schedule).
That creates a bias towards lines with high frequency thus with less variability and accounts for modelling the preference of the user for small variability
in the total travel time. 
Finally the transfer time 
function $t_{T}: T \rightarrow \mathbb{R}^{+}$ quantifies for each transfer $e^T_{\ell_1,\ell_2,j}$ the time needed to change between lines
$\ell_1$ and $\ell_2$ at node $j$. 

In such description the temporality of the system is included through the weights. The choice not to model the system as a temporal graph is motivated by the fact that in urban public transportation systems the total travel time is subject to variability and this factor matters considerably for the user when deciding to opt for public transportation service. 

\subsection{Uncovering efficient transportation connections}

The previously defined public transportation graph $G=(V, E, t_E, T, t_T)$ is used to calculate shortest time paths between stops. In the multi-edge representation a path is defined as a sequence of edges \footnote{ In the current paragraph, to simplify notations, we do not index edges by node names.} $P_E={\{e^{\ell_{i_1}},e^{\ell_{i_2}},...,e^{\ell_{i_n}}\}}_{o,d}$  connecting an origin node $o$ to a destination node $d$ through a sequence of  consecutive trips made on $n$ lines, ${\ell_{i_1},\ell_{i_2},...,\ell_{i_n}}$. Considering also the sequence of corresponding transfers between lines $P_T={\{e^T_{\ell_{i_1},\ell_{i_2}},e^T_{\ell_{i_2},\ell_{i_3}}...,e^T_{\ell_{i_{n-1}},\ell_{i_n}}\}}_{o,d}$ the shortest time paths between origin and destination are taken as the smallest durations measured among the different alternative paths. Each time length $L_{P}=\sum_{j=1}^{n}t_E(e^{\ell_{i_{j}}})+\sum_{j=1}^{n-1} t_{T}(e^T_{\ell_{i_j},\ell_{i_{j+1}}})$ is defined as the sum of the average time needed to wait, travel and transfer between lines. To calculate shortest time paths between any stops we adapt the Dijkstra algorithm \cite{citeulike:180287} to multi-edge graphs. While the original version of the algorithm computes the minimal distance between 
any origin $o$ and destination $d$ nodes by considering the sum of link weights, the modified version accounts for the fact that not only the link weights have to be taken into consideration but also the transfer time, i.e. the cost to change between links (see SI, section 3 for more details).

To consider the preference of users to change lines a limited number of times we allow at most two transfers in a single path, i.e., we limit $n\leq 3$. After computing the shortest paths between all nodes in the graph we characterise the distribution of shortest travelling times between all nodes whose physical distance falls within a specific range. Privileged connections at a specific distance range - timely speaking - can be identified by selecting the lower part of the travel time distribution (i.e. the best times for a given distance), which gives the best connections at a given distance relatively to the public transportation system considered.

\subsection{Implementation of the user-based representation}

The methodology presented above has been implemented to be applicable to transportation network informations given in GTFS format 
\footnote{\url{https://developers.google.com/transit/gtfs/reference}} where trips, routes, travelling times, frequencies and transfer times 
are recorded for each service line and station in the transportation system (for further details see SI, section 1). This type of data is openly 
accessible 
for several cities worldwide, from which here we consider municipal areas in France. We build the P-space multi-edge representations of larger Paris,
Strasbourg, Nantes ,and Toulouse by extracting the relevant information from corresponding GTFS datasets.
This procedure starts by identifying a 
period of $N_w=4$ weeks such that the total number of trips per day presents only weak fluctuations. Here, 
we are interested in trips planned between $h1=7am$ and $h2=10am$ 
(note that the choice of $N_w$, $h1$, and $h2$ are adjustable parameters). 
 Such choice for the time window of study is motivated both because we focus on 
morning commuting patterns, and because during this time interval the frequency of services is considerably higher with respect to 
the rest of the day. Typical line frequencies and trip durations are then defined
as their averages over the selected time window over the four weeks.

Finally, building on the multi-edge P-space representation and the estimation of the typical times and frequencies, we compute the typical
shortest time paths between any origin and destination in the city. The implementation of this methodology is available online\footnote{\url{ https://github.com/lalessan/user_basedPT}} and requires as input any dataset in GTFS format and parameters summarised in the SI.

\section{Illustration: fingerprints of public transportation networks}
We demonstrate one possible use of our framework through the examples of the PT systems of larger Paris, Strasbourg, Nantes, and Toulouse. 
After selecting privileged connections, we apply non-negative matrix factorisation to the graph of the privileged connections to identify underlying patterns of privileged connections. 
Finally, we compare our findings with independent measures of commuting patterns, which allow us to give an estimation about the efficiency of
the PT systems. 

\subsection{Selection of efficient connections}

\begin{figure}[htbp!] \centering
  \includegraphics[width=\textwidth, keepaspectratio]{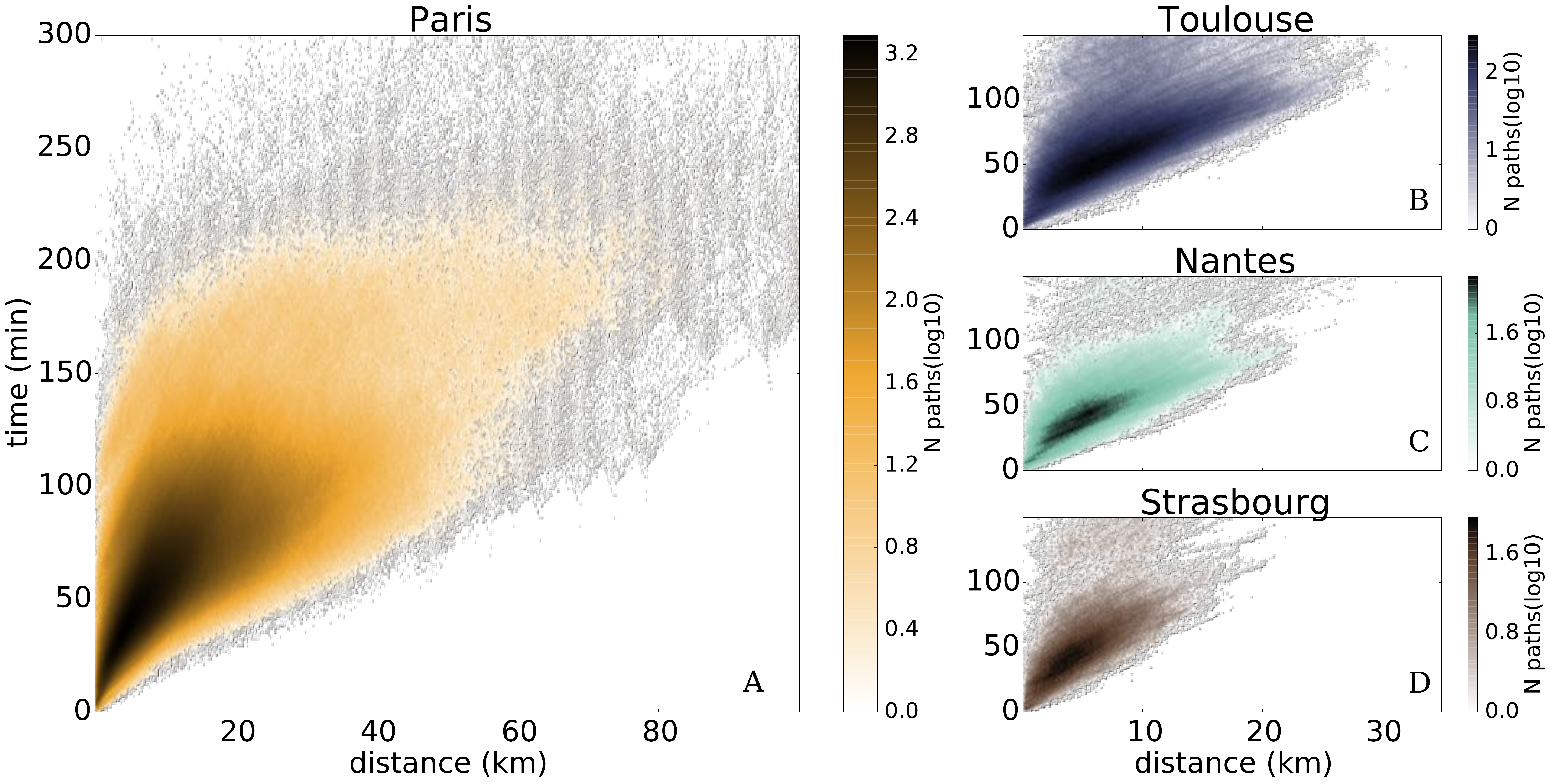}
\caption{ \textbf{Scatterplot of time versus physical distance associated to shortest time paths for each origin-destination pair.} The points are coloured
according to the number of points in the area considered. Scatterplots
are shown for the cities of Paris (A), Toulouse (B), Nantes (C), and Strasbourg (D). We considered distance ranges between $0$ and $d_{max}$ 
divided into $N_d$ number of bins with equal size $\Delta d_{bin}=100$ meters. In a similar way we divided the considered time range of $3$ hours 
into $N_t$ number of time bins of size $\Delta t_{bin}= 1$ minute. Colours indicate the logarithm of the number of origin-destination pairs in a
given range time-distance bin.}
\label{fig:2DHist}
\end{figure}

We used the method previously presented to compute the shortest time paths for each origin-destination pairs of the transportation systems of bus, 
train and metro (when possible). Based on these calculations we built a time-distance map, which assigns 
the physical distance $d(o,d)$ and the shortest time path length $\Delta t(o,d)$ to each origin ($o$) - destination ($d$) pair.
This time-distance map was drawn as a heat-map in Fig.~\ref{fig:2DHist} for Paris and the other investigated cities,
and can be used to identify patterns of privileged connections.

\begin{figure}[htbp!] \centering
 \includegraphics[width=0.8\textwidth, keepaspectratio]{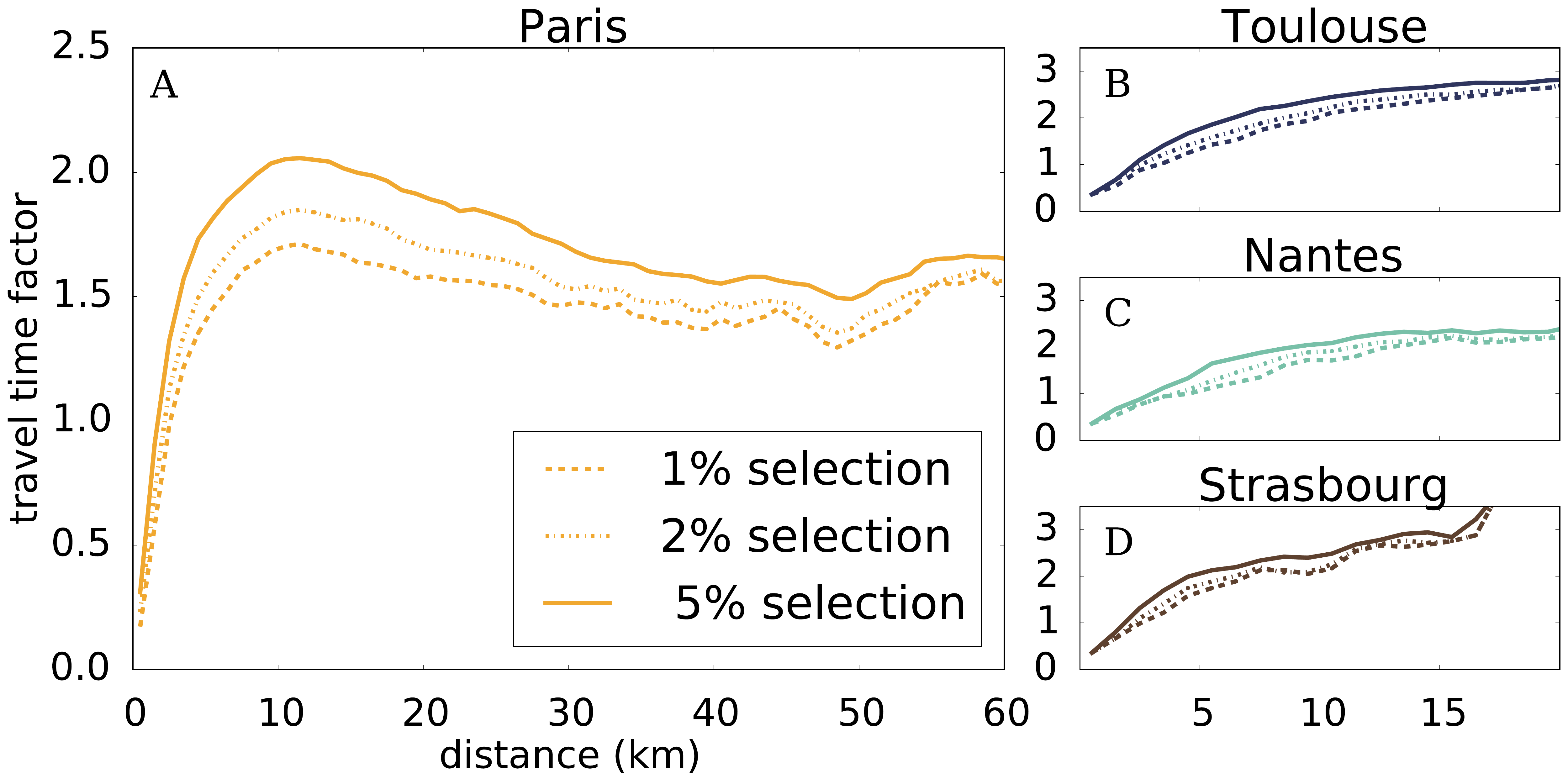}
\caption{ \textbf{Travel time factors with respect to distance travelled.} The factors have been computed
using the $1\%,2\%$ and $5\%$ lower part of the time distribution for each distance travelled by public transportation for the following cities (including
their surrounding areas) (A) Paris, (B) Toulouse, (C) Nantes, and (D) Strasbourg.}
\label{fig:selection}
\end{figure}

In order to focus on the most efficient connections with respect to the public transportation system of the city considered,
we selected the trips responsible for the $1\%$ lower part of the time distributions for each distance. To estimate, how these connections
among the best at the urban agglomeration level compare to travels by car for the same distances, we computed the travel time factor, 
i.e. the ratio between the travel time taken by public transportation and by car, for each urban agglomeration, as explained below. 
After building the histogram of shortest time paths for every distance bin, we compared the travel time of selected
paths with the travel time needed to cover the same distance by car. Car commuting times were extracted from the French 2008 Enqu\^{e}te 
Nationale Transports et D\'{e}placements 2007-2008 dataset \cite{dataset:Enquete2008} describing the global mobility of people living in France. 
To collect this data individuals were asked how far (with resolution of 1 km), how long (with resolution of 1 minute), and by which transportation
mean they travel every day. Based on this dataset we computed the median of the travel time distributions at each distance using the entire sample 
to measure the typical time needed to commute to a particular distance by car. 
Similarly, we calculated the medians of the best $1,2,5\%$ of the time distribution at each distance  (i.e. shortest times for a given distance) 
travelled by public transportation. 
This enables to compute the travel time factor as 
displayed in Fig.\ref{fig:selection} for different selections of the best times taken by public transportation. 
 By selecting the  best connections responsible for the $1\%$ lower part of the time distributions for each distance, in Paris agglomeration,
 we consider trips whose duration is at most $1.71$ times the time needed by car. This is in agreement with the travel time factor tolerated by
 users \cite{asperges2007determinants}. Studies have revealed that commuters typically consider to take public transportation if the travel
 time factor does not exceed 1.6 \cite{asperges2007determinants}. For the other agglomerations studied, the travel time factor goes above this value
 for distances travelled greater than $5km$. We remark that while in Paris the travel time factor tend to saturate at large distance meaning that
 efficient connections exist also at the inter-city level, this is
not conspicuous for the other cities (see Fig.\ref{fig:selection}.b-d). Public transportation seems to
provide an efficient alternative to car mainly for short trips. 
 
Note that in the histograms and travel time factor calculations we do not use the best absolute time to travel a given distance by PT but
a time estimation of a user arriving blindly at a stop in order to take into account the preferences of users for path with small variability 
in time. 
In addition, the time travelled by car for each distance is taken from the data in the whole country and may thus be underestimated for cities 
subject to urban congestion trafic. These two points may lead to an overestimation of time travel factors.
Thus, here, the travel time
factor cannot be used directly as a criterion to select the best connections but only gives a common metric to look at the different public transportation
systems.

\subsection{Pattern extraction}

\begin{figure*}[htbp!] \centering
\includegraphics[width=.9\textwidth]{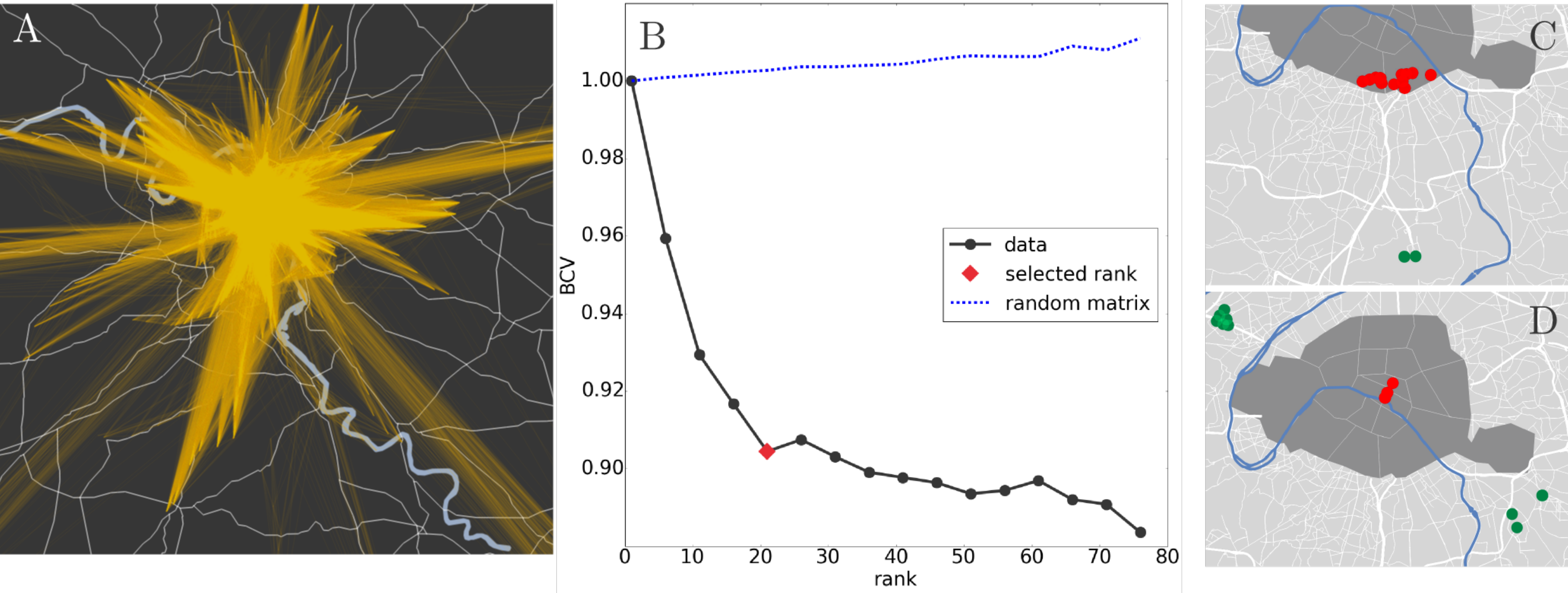}
\caption{\small{ \textbf{Pattern detection using the multi-edge P-space representation.} (A) Geographic representation of graph $G_{SP}$, 
where links correspond to the 1\% best shortest paths of the whole public transportation network. (B) 
The normalised BiCross validation error computed for the adjacency matrix $X_{SP}(10 Km,11 Km)$ (grey full line) of the same graph,
for the associated random matrix $X_{SPrandom}(10 Km,11 Km)$ (dashed line). The selected number of structures $k_{s}$ is assigned by a red rhombus.
(C,D) Two of the structures revealed in the PT system of Paris. Green dots are ingoing, while red dots are outgoing affiliated.}}
\label{fig:patt}
\end{figure*}

The goal of this section is to understand whether the identified set of privileged connections reveals any higher order meaningful patterns in 
the design of the transportation system that will help to characterise the network and to better  capture its underlying structure. We expect
that some stops, like stations located in residential neighbourhoods, may have similar connectivity patterns to the rest of the network e.g.
to the city centre or to working areas. In order to identify such patterns, we first built an undirected, unweighted graph
$G_{SP}=(V_{SP},E_{SP})$, where $V_{SP} \subset V$ and $E_{SP}$ is a set of edges linking origin-destination locations identified as being 
well connected (i.e. in the 1\% selection). Such a graph is displayed for Paris in Fig.~\ref{fig:patt}A. Then, to compare commuters travelling at particular distances we analysed subgraphs
$G_{SP}(d_1,d_2)$ of $G_{SP}$, where edges join stops at particular distances $d$ such that  $d_1< d \leq d_2$. Each subgraph $G_{SP}(d_1,d_2)$ 
is represented by an adjacency matrix $X_{SP}(d_1,d_2)$. For the city of Paris we considered intervals for distances with resolution
of $d_2-d_1=1$ kilometer.  For the other cities we considered distances with resolution
of $d_2-d_1=5$ kilometers as the transportation network is much less dense spatially for latter cases (see SI Fig.1). 

Both cohesive, and bipartite structures can exist in these subgraphs. The cohesive structures  would correspond to sets
of stations well connected between themselves while the bipartite structures would single out two groups of stops such that many connections 
exist between the two groups. The connections may not be direct but should have durations comparable to the average time taken by car by definition.
The likelihood of having a link between any two stations can be expressed in term of the possible links of these stations to the same structures.
Formally, it means we can express each term of the adjacency matrix representing $G_{SP}$ as $X_{SP}(i,j)=\sum_{k}W_{ik}H_{kj}$, where
 $W_{ik}$ is a  value quantifying the ingoing membership of node $i$ to structure $k$ and $H_{kj}$ a value quantifying the outgoing affiliation 
 of the node $j$ to the structure. In order to find matrices  $\mathbf{W}$ and  $\mathbf{H}$, we performed matrix factorisation, minimising
 numerically the distance $\| \mathbf{X} - \mathbf{WH} \|_F^2$, where $\|\mathbf{X}\|_F$ is the Frobenius norm of
 matrix $\mathbf{X}$ (for further details see SI,section 2). Matrix factorisation was shown to be successful in detecting communities and
 richer structures in graphs \cite{paper:NMFCommunityDetection, paper:NMFCommunityDetection2, paper:NMFCommunityDetection3, paper:NMFCommunityDetection4, paper:NMFCommunityDetection5, paper:NMFCommunityDetection6, paper:NMFCommunityDetection7}.

The number of structures to be detected was determined by the Bi-Cross validation (BiCv) approach proposed in \cite{paper:BiCrossValidation} based 
on cross-validation, a common machine learning model validation technique. This consists of measuring an error, called $BCV$ here, between 
an estimation of left out entries using a low rank approximation of the retained data and the actual left out entries. This error is decreasing
with respect to the number of structures extracted toward a minimum that indicates how many structures are representative of the subgraphs 
(more details in the SI, section 4). When the network is close to random, the error does not show such a behaviour. Thus, to identify whether there is a
lack of structures in subgraphs, we compared the $BCV$ error with the one obtained for the corresponding null models (Fig.~\ref{fig:patt}B)
defined for each adjacency matrix $X_{SP}(d1,d2)$ as its corresponding random matrices $X_{SPrandom}(d1,d2)$ with the same size and density. 
An example of the behaviour of such a quantity for Paris public transportation network is displayed on Fig. \ref{fig:patt}B (for other cities see SI, Fig.1).
This quantity was computed for each subgraph, i.e. subsets of shortest paths at different distance intervals. This guided us on how many structures
characterise each system at each range of distance. For some distance ranges and cities, the evolution of $BCV$ is close to the random case
meaning that one does not detect any strong attempt to link preferentially some areas at the considered range of distance (see SI, Fig.2). 
In several cases bipartite structures can be found, like the two examples in Fig.\ref{fig:patt}C and D detected in the Paris network. 
The bipartite structures can be assimilated to strategical areas that are particularly well connected by PT. For example the structure shown in figure
\ref{fig:patt}C, connects stops located around Paris Orly airport to stops located at the border of Paris central area. In figure \ref{fig:patt}D,
the structure reveals the existence of privileged connections the Nanterre and Creteil areas in one side (both with high employment density) \footnote{\url{http://insee.fr/fr/themes/document.asp?reg_id=20&ref_id=20718\&page=alapage/alap417/alap417_carte.htm\#carte1}},
with Paris
center on the other side. 
As these structures are latent patterns extracted from the networks of priviledged connections, we consider them as the privileged origin-destination
patterns representative of the transportation systems.

\subsection{Network efficiency: pattern analysis from the commuter point of view}

To estimate how well the different public transportation networks are devoted to answer the needs of commuters, we compared the identified 
privileged origin-destination patterns to the flows of commuters. To do so, we used data of the 2010 French census
\cite{dataset:MobPro2010} including origin-destination commuter flows per transportation mean at the level of the municipality for
the large areas of Strasbourg, Toulouse and Nantes and at the level of the municipal arrondissement (neighbourhood) for
Paris agglomeration. Using this dataset we compared the detected privileged origin-destination patterns to the commuting patterns
by car and PT. We only consider inter-municipality trips for the comparison as the resolution provided for
the commuter dataset is given at the municipality level. The number of intra-city trips is provided in SI Tab.S4.

To draw a comparison, we first built the PT structural pattern network $G_C=(V_C,E_C)$ of each urban agglomeration as an unweighted, undirected graph. 
The set of nodes $V_C$ is defined as municipalities and a link $(a,b)\in E_C$ between municipalities $a$ and $b$ exists if
at least one stop located in $a$ and one stop in located $b$ appear in each side of a detected bipartite structure. In other words, the
structural pattern networks are composed of links between municipalities presumably well connected by public transportation.
At the same time, exploiting census data, we built a commuter flow network for each city and its surrounding area, as a weighted, directed 
graph $G_{com}^{TM}=(V_{com}^M,E_{com}^{TM}, W_{com}^M)$.  Here $V_{com}^{TM}$ is the set of municipalities, and a 
link $(a,b) \in E_{com}^M$ with weight $w_{ab}$ represents the flow of individuals commuting from $a$ to $b$ by mean 
$TM$ (either PT or car). Then we compared the structural pattern  graph with the commuter flow graphs both of the car and the PT
of each urban agglomeration by 
computing a weighted Jaccard index $s$ between the sets of links associated to each graph. This weighted index is defined as the sum
of the flow graph weights of the links in common between the two graphs - structural and flow by the selected transportation mean - divided by the total flow for the transportation mean considered. The weighted Jaccard index is calculated as follows: 
$s_{TM}=\frac{\sum_{(a,b) \in E_C \cap E_{com}^{TM}}{w_{ab}}}{\sum_{(a,b) \in E_C \cup E_{com}^{TM}}w_{ab}}$ both for $TM=car$ and $TM=PT$. It
 represents the fraction of commuters using respectively car and PT, who have access to privileged PT 
connections (i.e. for which there exists a link corresponding to their commute in the PT structural pattern).

 \begin{figure*}[htbp!]
\centering
\includegraphics[width=.9\textwidth]{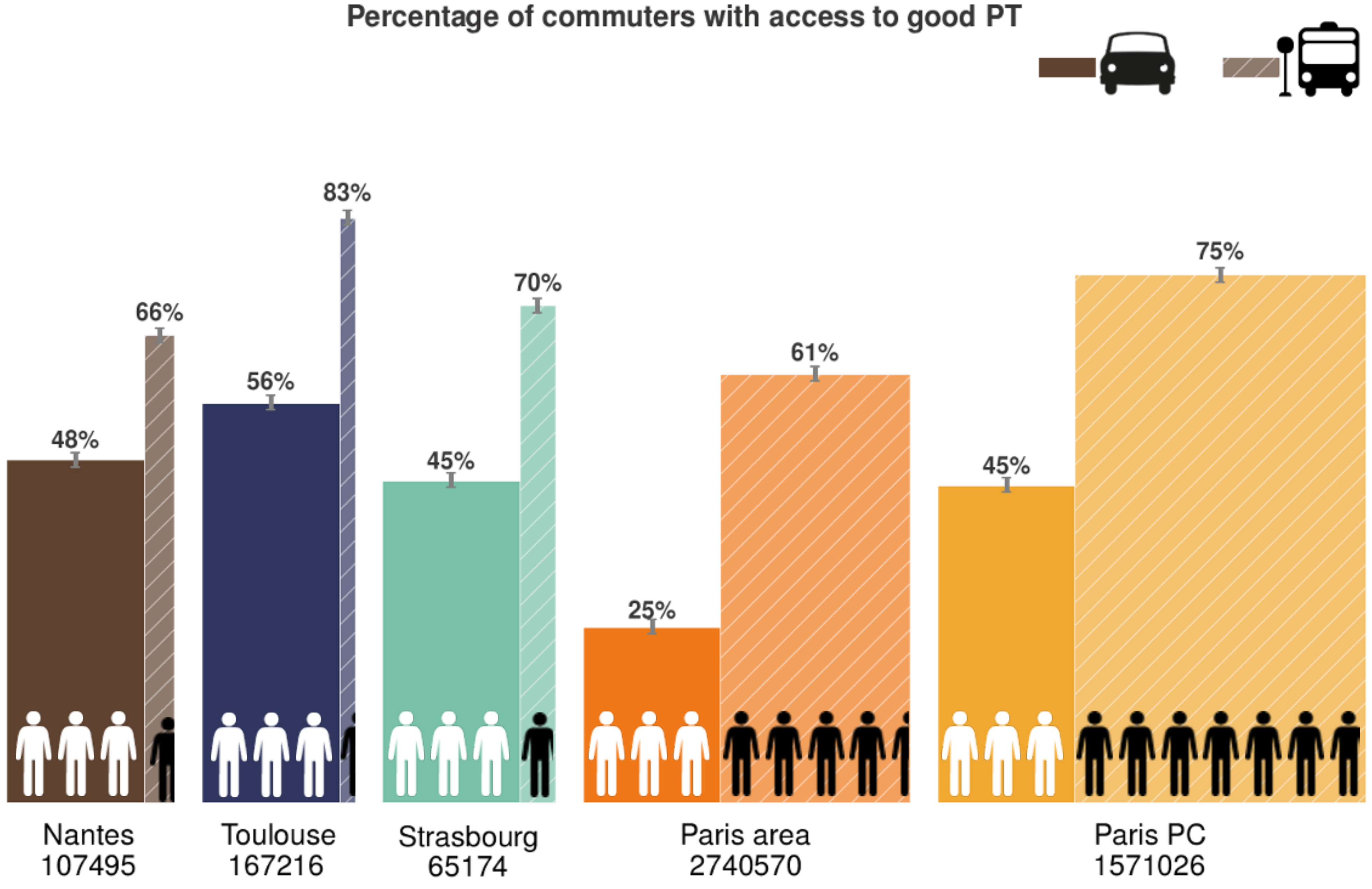}
\caption{\textbf{Measures of PT network efficiency in French municipal areas.} For each of the urban agglomerations considered 
(Toulouse, Nantes, Strasbourg, Paris area, and Paris Petite Couronne), the bar chart's height indicates the weighted Jaccard index $s_{TM}$
between the commuter flow network $G_{com}^{TM}$ and the PT structural pattern network $G_C$. Plain bars refer to the case where the transportation
mean used is the $TM=car$, while dashed bars  refer to public transportation systems $TM=PT$. The width of plain bars are set to be equal for all the cities considered. The width of the dashed bar filled
is set proportionally. Three stick men represent the population having access to privileged connections (defined as previously) and using a car 
to commute such that the ratio between the number of black and white stick men is the same as the ratio between the total number of commuters travelling by 
car and by PT for each of the urban agglomerations considered. The total number of commuters (using car or PT) in each city is indicated below each bar. Error bars on the top of the
bars are obtained by repeating the methodology 100 times, with different random matrices initialising NMF. The smallness of the error bars shows the
robustness of the pattern detection.
}
\label{PTvsCar}
\end{figure*}
Bar charts of Fig.~\ref{PTvsCar} show comparisons among weighted Jaccard indices measured by considering the car commuting flow $s_{car}$ and the 
PT commuting flow $s_{PT}$ for several urban agglomerations.
A significant difference between the commuting practice in Paris agglomeration and other urban areas is observed. 
For Paris urban agglomeration, the flow of inter-municipality commuters choosing PT is larger than that of people commuting by car, in contrast
to the other investigated cities. This may be partly explained by a travel time factor, which increases above the tolerated value for Toulouse, Nantes and Strasbourg (see Fig.~\ref{fig:selection}). 
Besides, Figure \ref{PTvsCar} indicates that the fraction of  commuters having access to privileged connections
(defined as previously) and using the PT systems is larger than the fraction of them using the car for all urban agglomerations studied. This
comforts our definition of privileged connections based on commuting time with little variability and a limited transfer number. This
 corroborates the strong role of the latter factors in the decision making to use PT or car.
Furthermore, we observe that in the larger Paris area only $25\%$ of car commuters have access to privileged transportation connection. Instead, 
in other cities, although more than $48\%$ of car drivers have access to presumably privileged connections, they still
commute by car. In particular, in Toulouse a large percentage of commuters have access to good services according to the criteria introduced here, as there is large
overlap between privileged connections and both PT ($83\%$) and car ($56\%$) commuting flows, however, there is still a non-negligible amount 
of people commuting by car. We distinguish two main trends in commuting: there are cities where a large part of the population
tend to do inter-municipality trips by car disregarding the quality of PT services based on the aforementioned factors, examples are Nantes, Toulouse, and Strasbourg. On the other hand, 
in Paris and its agglomeration, according to the metrics introduced, there is a good agreement between the needed and provided services of public transportations.

\section{Conclusion}

The framework presented here along with its implementation allows us to gain insight in the structure of public transportation networks from a
user point of view. It is detaching from the constraints imposed by the spatial embeddedness of this type of network and incorporates temporal 
variance of its schedule. It gives a way to characterise the public transportation system of different cities by identifying where some efforts
have been put, not only structurally, but also in terms of the frequency of the connections. As an illustration for the use of the presented 
methodology we quantified the efficiency of transportation systems regarding the flow of commuters, in other words, how well the transportation 
system answers the need of its users.

Several extensions of our methodology are possible. Parameters like the periods in focus, length of observations, number of transfers, etc.
can be tailored for other systems, while a further refinement is possible by considering needs of other types of users. Regarding the structures
extracted by the factorisation method, even though one sees that many of them are linked to commuting patterns, one can focus on their more precise 
interpretation in a future study. 

\section*{Acknowledgment}

This research was supported  by the Lagrange Project of the ISI Foundation funded and by the CRT Foundation by the FET Multiplex Project (EU-FET-317532) funded by the European Commission. LA is thankful for the DANTE Inria team and the ISI Foundation for funding her internship and for ENS de Lyon for the Amp\'ere Excellence fellowship. The funding bodies had no role in study design, data collection and analysis, preparation of the manuscript, or the decision to publish.


\newpage

\begin{center}
{\LARGE Supplementary Information for}\\[0.7cm]
{\Large \textbf{User-based representation of time-resolved multimodal public transportation networks}}\\[0.5cm]
{\large Laura Alessandretti, M\'arton Karsai and Laetitia Gauvin$^{*}$}\\[0.7cm]
{\small $^*$Corresponding author email: marton.karsai@ens-lyon.fr}\\[2cm]
\end{center}

\section{Data description}
With the aim of catching a comprehensive picture of the public transportation (PT) networks in French municipal areas we made use of datasets
provided by local public transportation companies. The characteristics of the datasets used for the different cities are listed in
Table S\ref{Table1}. 
Estimated timetable schedules for the public transport service are made publicly available online and 
frequently updated by the companies.

\renewcommand{\arraystretch}{1.3}
\begin{table}[h!]
\begin{tabular}{ l l l l l}
 \textbf{City} &  \textbf{Area} &  \textbf{Period} & \textbf{Companies}  \\
\hline
Paris & 47.96N-49.45N 1.15W-3.51W & Sep-Oct 2013 & \begin{tabular}{ll}RATP (Bus, Metro, Tram, RER) \\ SNCF (RER,Train)\end{tabular}\\
\hline
Toulouse& 43.43N- 43.74N 1.17W- 1.69W & Sep-Oct 2014 & \begin{tabular}{ll}Tiss\'{e}o (Bus, Tram, Metro)\\ SNCF (Train)\end{tabular}\\
\hline
Nantes& 47.12N- 47.32N 1.75W-1.34W & Jan 2015 & \begin{tabular}{ll} Semitain (Bus, Tram, Ferry)\\ SNCF (Train)\end{tabular}\\
\hline
Strasbourg& 48.46N- 48.68N 7.60W-7.83W & Jan 2015 &  \begin{tabular}{ll} CTS (Bus, Tram)\\ SNCF (Train)\end{tabular}\\
\hline

\end{tabular}
\caption{Table listing the main characteristics of the data used for each of the cities }

\label{Table1}
\end{table}

All datasets are provided in General Transit Feed Specification (GTFS) format \cite{sito:GTFS}. GTFS is a common format for PT schedules and
associated geographic information. It is composed of a series of text files: stops, routes, trips, and other schedule data. In particular, 
the following objects and associated attributes are of relevance to the purpose of this study:
\begin{itemize}
\item{\textbf{stop}: the physical location where a vehicle stops to pick up or drop off passengers. It is associated to a unique
\textit{stop\_id} and it has attributes \textit{stop\_name}, \textit{stop\_lat}, \textit{stop\_lon}, respectively the name and the geographic 
coordinates. (Example: 4025460, "PONT NEUF - QUAI DU LOUVRE", 48.858588, 2.340932)}

\item{\textbf{route}: a public transportation line (in the following we  refer to "line" or "route" as interchangeable terms) identified by a unique \textit{route\_ id}. It has attributes \textit{route\_ type}, identifying the type of vehicle, and \textit{route\_ name}. (Example : 831555, metro, "14")}

\item{\textbf{trip}: a journey of a vehicle, identified by a unique \textit{trip\_id}. It refers to the unique route of the actual line, and also to a set of dates indicating in which days of the year that trip is running. It is also associated to an ordered sequence of stops of the vehicle, and with the list of
arrival and departure time at each stop.\\ \\
Example: \hspace{100 pt}
\begin{small}
\begin{tabular}{ c c c c }
 trip\_id& stop\_id & arrival\_time & departure\_time \\
1013644000942075 & 4025388 & 16:10:00 & 16:10:00\\
& 4025390 & 16:11:00 & 16:11:00\\
& 4025392 & 16:12:00 & 16:12:00\\
& 4025393 & 16:13:00 & 16:13:00 \\
& ... & ... & ...\\
\end{tabular}
\end{small}
}
\end{itemize}

\renewcommand{\arraystretch}{1.3}
\begin{table}[h!]
\begin{tabular}{ c c c c c c}
 \textbf{Area} &  \textbf{Stops} & \textbf{Routes} & \textbf{Train stops} &  \textbf{Train routes} & \textbf{Tot stops after merging} \\
\hline
Paris & 11850 &  1058 & 494 & 169 & 5690 \\
\hline
Toulouse  & 1913 &  106 & 59 & 31 & 1920\\
\hline
Nantes & 3412 & 61 & 27 & 18 & 1038\\
\hline
Strasbourg & 1330 &  53 & 31 & 17 & 601\\
\hline
\end{tabular}
\caption{Table illustrating the main characteristics of the PT systems datasets. For each urban agglomeration (Area), we indicate the number of Bus, 
Metro, Tram and RER stops and routes before coarse graining (Stops, Routes), the number of train stops (Train stops) and routes (Train routes), the total number of stops after coarse-graining and merging the two datasets (Tot stops after merging).  }

\label{Table2}
\end{table}

\subsection{Coarse graining network stops}
To model the transportation network, it was necessary to coarse grain the data by grouping nearby stops together. 
Table S\ref{Table2} summarises the information contained in each of the datasets before and after coarse-graining.

\subsubsection{Paris}
The transportation system described in the RATP dataset contains 11850 stops. Some of these stops
 closely located to each other can be functionally  replaced by a single station via a careful merging method. 
 In order to merge stops, we used the information provided in the GTFS dataset. Data provides the list of stop pairs that are located at 
 a short distance from each other, allowing people to transfer walking, from one route to a different one in a given amount of time 
 (that is also given in the dataset). It is for example the case of main railway stations or big squares, where many stops are concentrated 
 in a relatively small area. We merged corresponding stops according to the information provided by the RATP company on possible transfers, as
 well as bus stops located in front of each other at the two opposite sides of the same road. After coarse graining, the total number of stops
 for the RATP dataset was reduced to 4596.

In the SNCF dataset, there is a total number of 494 suburban railway stations. It is necessary to identify stops/train stations present both in the SNCF 
and RATP datasets (i.e "Gare du Nord" is both a RER station and a metro stop). To do so we built a grid with a resolution of 0.25 Km and we identified
for each of the train stations the cell it belongs to. A train station was then identified by the closest RATP stops present in the actual cell
or in neighbouring cells otherwise. In the city centre, all the train stations were identified with RATP stops, while in the suburbs it was not
always the case.

\subsubsection{Nantes}
The Semitain dataset contains 3412 stops. It indicates for each stop whether it is part of a larger station complex
(stops that are located on the opposite side of a same road are considered part of a unique station). Using such information, it was
straightforward to merge close-by stops. After coarse graining, the network includes 1036 stops. The SNCF dataset was used to include
the train stations which are located in the area served by the Semitain company. Using the same method we used for Paris, we found
their corresponding stops in the Semitain dataset. 

\subsubsection{Toulouse}
The Tiss\'{e}o dataset contains 5694 stops. As in the case of Nantes, the Tiss\'{e}o dataset provides information on parent stations. We merged stops accordingly received 1913 stops in total. From the SNCF dataset, we selected 59 stops that located in the same area served by the Tiss\'{e}o company.

\subsubsection{Strasbourg}
The CTS dataset contains 1330 stops. Even if it does not provide information on parent station, we could merge stops based on their
\textit{stop\_id}. In fact, in this dataset all stops that are part of a larger station complex have the same name and in addition a
unique number (Example: stops \{DANTE\_01, DANTE\_02, DANTE\_03\} are part of a same large station complex). After coarse graining this way 595
stops were identified in the CTS dataset. From the SNCF dataset, we selected 31 stops that are located in the same area served by the CTS company.

\subsection{Choice of a representative day}
The datasets provide the schedule over several  months in normal situations (which means no perturbation due to traffic jams or to system breakdowns)
with a 1-minute resolution.
we do not consider exact travel time at a given departure time but an estimation of the time taken in a ``typical''  day. The description of a typical
day is given below.

In order to draw typical commuting times we first selected a window of $N_w=4$ consecutive weeks. A week $w_i=\{d_1,d_2,d_3,d_4,d_5\}$ is
 defined as a set of five consecutive days, from Monday to Friday. The separation week-end/week days is necessary as the
 system behaviour is different in these two cases. For every span of consecutive weeks $W=\{w_1,w_2,w_3,w_4\}$, we 
calculated the average daily number of trips $\langle Nt_{W} \rangle =\sum_{d \in W} Nt_{d}/D$. Here $D$ is the number of days 
($D=5 \times 4=20$), $Nt_d$ is the number of trips during day $d \in W$. Then, by looking at fluctuations from the average
$\sigma_W^2=\sum_d (Nt_d - \langle Nt_{W} \rangle )^2/D$, we selected the four weeks span $W$ for which $\sigma_W^2$ is the smallest. For each 
city the selected period is indicated in Table S\ref{Table1}.

The reason to select a span of time where the number of trips is not fluctuating is motivated by the need to work with meaningful averaged quantities. We are aware that the results of the illustration may not generalise, as they are relative to a specific selected period of time. Future work could include a comparison to the system behaviour during weekends, and at different times of the year.

For the purpose of this work, as we aimed at comparing our results with the flux of commuters, we limited the analysis to the 7-10am time interval. Indeed, as a first step, we selected all trips occurring between $h1=7 am$ and $h2=10 am$ within the selected period. Further work could include the study of the system evolution at different times of the day. 

As a second step, we calculated for each route $\ell_k$ and each day $d\in W$ the total number $Nt_{\ell_k,d}$ of trips $tr$ occurring on day $d$ between 7 and 10 am and computed its average over the four selected weeks $\langle Nt_{\ell_k} \rangle =\sum_{d \in W} {Nt_{\ell_k,d}}/D$. In this way, we received the average frequency $f_{\ell_k}=\langle Nt_{\ell_k} \rangle /3h$ (3h is the length of the time interval) in the selected period for each metro, bus or train line. Also, we computed in equivalent way, the average duration of a trip between any two stops $i$ and $j$ along line $\ell_k$: $\langle \Delta t_{ij}^{\ell_k} \rangle=\sum_{tr}( \Delta t_{ij}^{\ell_k})/ \sum_{d \in W} {Nt_{\ell_k,d}} $ considering all selected trips $tr$.

\subsection{The INSEE datasets}
In order to analyse commuting patterns, we gathered two datasets of the French Institute of Statistics (INSEE): 
the \emph{Enqu\^ete Nationale Transports et D\'eplacements 2007-2008} \cite{site:enquete_transport}
used for computing the commuting travelling times, and the 2010 French census (\emph{Recensement de la population 2010}) \cite{dataset:MobPro2010}
to extract origin-destination commuting patterns. 

We used the  file ``Q\_ind\_lieu\_teg.csv" of the \textbf{first dataset} providing for each individual several informations about 
their daily journey to work/school. We estimated the average time needed to commute a specific distance by car by 
scanning over the following variables \emph{V1\_BTRAVDIST}, i.e. the distance covered daily 
(resolution 1Km), \emph{V1\_BTRAVTEMPSA} i.e. the time needed to cover such distance (5 minutes resolution), 
and \emph{V1\_BTRAVMOYEN1S}, i.e. the transportation mean used. The time computed for a given distance is the time average over the trips with the same distance
and travelled by car.

The flow of commuters for each origin-destination trip was estimated using the file
``FD\_MOBPRO\_2010.txt" of the \textbf{second dataset}, in  which each line provides several variables related to an individual interviewed.
In particular, the following variables were needed: \emph{COMMUNE} and \emph{ARM}, respectively indicating the INSEE code associated to the 
municipality and the arrondissement (available only for central Paris) where the individual interviewed lives,
 \emph{DCLT} the INSEE code indicating the municipality and the neighbourhood (only for Paris) of work, and \emph{TRANS} referring to
the transportation mean used to commute (either by foot, two-wheeler, car/camion/van, PT). We also considered the variable \emph{IPOND} to
take into account that, because not every single citizen is interviewed for the census, each individual has a statistical weight to infer a representative
behaviour. 
Tables S\ref{Table3} and S\ref{Table4} provide an overview on the data for each of the urban agglomerations considered for this study.

\begin{table}[h!]
\begin{center}
\begin{tabular}{c c c c c c}
 \textbf{Area} &  \textbf{Mun} & \textbf{O-D pairs} & \textbf{Tot comm} & \textbf{Car comm} & \textbf{PT comm} \\
\hline
Paris & 460& 61897 & 4321011 & 1542640 & 2017768 
\\
\hline
Toulouse  & 89 &  2319 & 363679 & 249642 & 57269
\\
\hline
Strasbourg & 57 & 618 & 170337 & 92275 & 36576
\\
\hline
Nantes &	26 & 524 & 225026 & 143441 & 45455 
 \\
\hline
\end{tabular}
\caption{ For each one of the urban areas considered (Area), the table provides with the number of municipalities considered (Mun), 
the number of origin-destination pairs travelled by commuters (O-D pairs), the total number of commuters (Tot comm), the number of commuters
travelling by car (Car comm), the number of commuters travelling by PT (PT comm). }
\label{Table3}
\end{center}
\end{table}

\begin{table}[h!]
\begin{center}
\begin{tabular}{c c c c c }
 \textbf{Area}& \textbf{IC comm} & \textbf{IC car comm} & \textbf{IC PT comm}  \\
\hline
Paris & 1369535 & 390105  & 429735
\\
\hline
Toulouse  &  187797 & 99142 & 40553
\\
\hline
Strasbourg & 99041 & 40579 & 23098
\\
\hline
Nantes &111434 & 55444 & 25957
 \\
\hline
\end{tabular}
\caption{ For each one of the urban areas considered (Area), the table provides with the total number of intra-city commuters (IC comm), the number
of intra-city commuters using the car (IC car comm),  the number of intra-city commuters using PT (IC PT comm).}

\label{Table4}
\end{center}
\end{table}

\subsection{Matching the INSEE datasets and the PT datasets}
In order to establish a comparison between the commuting patterns and the efficient connections of the transportation systems, 
we matched the INSEE dataset with the PT data by associating to each of the stops in the PT data its corresponding municipality (or 
neighbourhood in the case of Paris). We made use of the Google Maps API \cite{site:googleAPI} 
to assign to the latitude-longitude coordinates of each PT stop its corresponding address. Then, we matched the municipality to its 
corresponding INSEE code via the file \emph{Base communale des aires urbaines 2010} provided by INSEE. \cite{site:base_communes}

\section{Structure detection with Non-Negative Matrix Factorisation}
In this section, we explain non-negative factorisation was achieved in order to extract structures from the transportation system dataset.
\subsubsection*{Algorithm}
Aiming at minimising the Euclidean distance loss function between the original matrix and the factorized one, we implemented the standard
multiplicative rule developed by Lee and Seung in \cite{paper:NMFAlgo}:
\[
\mathbf{H}_{ci} \leftarrow \mathbf{H}_{ci}\frac{(\mathbf{W}^T \mathbf{V})_{ci}}{(\mathbf{W}^T\mathbf{W}\mathbf{H})_{ci}} \; \; \; \;\;\;\;\;\;\;\;
\mathbf{W}_{ic}\leftarrow\mathbf{W}_{ic}\frac{(\mathbf{V}\mathbf{H}^T)_{ic}}{(\mathbf{W} \mathbf{H} \mathbf{H^T})_{ci}}
\]

\subsubsection*{Initialisation}

The NMF algorithm may not converge to the same solution at each run, depending on the initial conditions. To address this problem we initialise the matrices $\mathbf{W}$ and $\mathbf{H}$ randomly and run the algorithm 500 times. At each iteration we compute the divergence $\| \mathbf{V} - \mathbf{WH} \|_F^2$ and we select the iteration for which the error was minimal.

In the present case, the algorithm turns out to be stable and the results are robust for large networks, 
future development of this work could however include the study of a consensus clustering procedure. Consensus clustering 
is the problem of reconciling clustering information about the same data set coming from different runs of the same algorithm. For NMF,
some efforts have been done in this directions \cite{paper:NMFConsensus}, however, as the result of the clustering is described through 
two different matrices and the partitioning is soft, the problem is not trivial to solve.

\subsubsection*{Soft/Hard partitioning}

The results of NMF provide a soft clustering of the stops to the structures. Such information is included in matrices $\mathbf{W}$ and  $\mathbf{H}$. For a given node $i$ and a given structure $k$, $W_{ik}$ is the out-going affiliation of node $i$ to structure $k$, while $H_{ki}$ is the in-going affiliation.
As the original matrix can be very sparse, and the NMF algorithm can hardly produce zero-values, many of the values in $\mathbf{W}$ and $\mathbf{H}$ are positive but very close to zero. In order to overcome this problem and to make sure we are capturing only the most relevant information, we applied a method to binarise the matrices $\mathbf{W}$ and $\mathbf{H}$ as follows: For each structure $c$, vectors $\mathbf{H_{c}}$ and $\mathbf{W}^T_{c}$ contain respectively the in-going and out-going affiliation of each node $i \in V$ to the structure $c$.  With the goal of selecting only nodes that are strongly affiliated to $c$, we applied k-means clustering on these two vectors. k-means clustering partitions the $|V|$ affiliation values into $k$ clusters.
By choosing $k=2$ for each of the structures $c$ we distinguished a subset of not-affiliated nodes, whose affiliation value 
was very small, and a subset of affiliated nodes, whose affiliation value was significantly different from zero. Using this partitioning 
we defined a binary matrix $\mathbf{H'}$ such that $\mathbf{H'_{ci}}=1$ if node $i$ is in-going affiliated to community $c$ and
$\mathbf{H'_{ci}}=0$ if it not. In the same way, we define $\mathbf{W'}$, for the out-going affiliation.

\section{The modified Dijkstra algorithm}
As we have discussed in the main text we devised a modified version of the Dijkstra algorithm allowing to compute shortest paths in a weighted, labeled-edge graph. Below we present the pseudo code of the modified algorithm. The algorithm requires:
\begin{itemize}
\item{A graph $G=(V, E, t_E, T, t_T)$ with vertex set
$V$ with cardinality $N$, edge set $E$ with weight function $t_E$, and set of transfers $T$ with weight function
$t_T$}
\item{ A cut-off $L_{max}$(the maximal number of line changes allowed)}
\end{itemize} 
The algorithm returns:
\begin{itemize}
\item{An array $dist$ of length $N-1$, where $dist[u]$ is the shortest path length between nodes $s$ and $u$}
\item{an array $ \Pi_{node}$ of length $N-1$, where $ p=\Pi_{node}[u]$ is the \emph{parent node} of node $u$, that precedes it in the shortest path between the source $s$ and $u$ itself}
\item{the array of \emph{parent edges} $ \Pi_{node}$, of length $N-1$, where $ \Pi_{edge}[u]$ is the edge connecting $u$ and its parent node $p$ in the shortest path connecting $u$ and the source $s$}
\end{itemize}
In the pseudo-code, the following notations are introduced: $lenPath$ assigns to each vertex $v$ the number of edges to reach source
$s$, $Q$ is a min-priority queue initialised with all nodes in $V_G$, where priority is given to nodes that are at shortest distance 
from the source $s$, $EXTRACT-MINQ$ is the operation of selecting and removing the node with highest priority from $Q$, $e^{\ell_k}_{uv}$ is 
an edge in $E$ connecting nodes $u$ and $v$ via line $\ell_k$, and $u$ is a neighbour of $v$ if at least one of such connections exists and $e_m$ is the edge connecting two nodes in the fastest way, also taking into account possible line transfers when coming from an other node, $t_m$ is the associated time.

\begin{algorithm}[H]
\caption{Dijkstra on P-Space Multiplex network\label{alg:dijk8}}

\begin{codebox}
    \li \For each vertex $v \in \id{V_G}$
    \li \Do
            $\at{dist}{v} \gets \infty$
    \li     $\at{\Pi_{node}}{v} \gets \const{NIL}$
    \li		$\at{\Pi_{edge}}{v} \gets \const{NIL}$
    \li		$\at{lenPath}{v} \gets 0$
        \End
    \li $\at{dist}{s} \gets 0$

    \liempty

     $Q \gets \id{V_G}$

    \li \While $Q \neq \emptyset$
    \li \Do
            $u \gets \proc{Extract-Min}{Q}$

    \li     \For each $v$ in neighbors $u$:
            \Do
    \li     \If $\at {\Pi_{edge}}{v}==NIL$:
    \li     \Then  $t_m,e_m=min,argmin(t_E(e^{\ell_k}_{uv}))$

    	\li		\Else
    	\li			$t_m,e_m=min,argmin(t_E(e^{\ell_k}_{uv})+t_T(\Pi_{edge}(u), e^{\ell_k}_{uv}))$

    			\End
    			
    	\li
    	\liempty
                \If $\at{dist}{v} > \at{dist}{u} + t_m$ AND $\at{lenPath}{u}+1 <=L_{max}$
    \li         \Then
                    $\at{dist}{v} \gets \at{dist}{u} + t_m$
    \li             $\at{\Pi_{node}}{v} \gets u$
    \li 				$\at{\Pi_{edge}}{v} \gets e_m $
    \li 				$\at{lenPath}{v}= \at{lenPath}{u}+1 $
    \liempty
                \End
            \End
        \End

        \Return $dist, \Pi_{node}, \Pi_{edge}$
\end{codebox}

\end{algorithm}

\section{Pattern detection for Strasbourg, Nantes, and Toulouse}
For the urban agglomerations of Strasbourg, Nantes, and Toulouse we detected structural patterns by considering intervals for distances with 
resolution of $d_2-d_1=5$ kilometers. An example  of structure detected for each city is shown in Fig.SS\ref{SchemaPaths}. 
For an interval range $(d_1,d_2)=(5,10) km$, both for Strasbourg and Nantes, we observed that the BiCross validation error computed for the adjacency matrix $X_{SP}(5 Km,10 Km)$ is similar to the BiCross validation error of the associated random matrix $X_{SPrandom}(0 Km,5 Km)$ (Fig.SS\ref{noStructure}). This suggests that there is a
lack of structure in the subgraph $G_{SP}(5 Km,10 Km)$.

\begin{figure*}[htbp!]
\centering
\includegraphics[width=.9\textwidth]{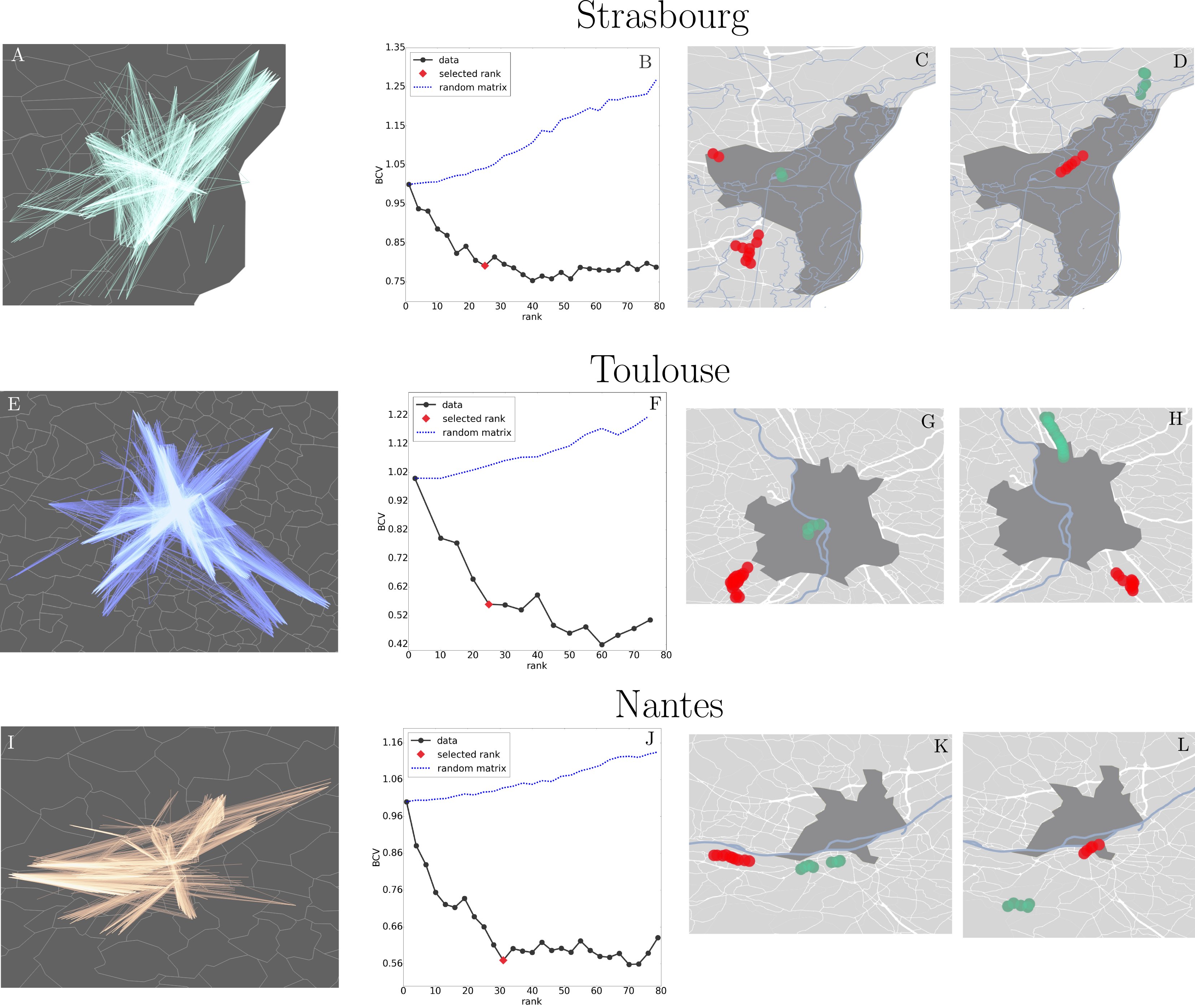}
\caption{\small{ \textbf{Pattern detection using the multi-edge P-space representation.} For Strasbourg, Toulouse, and Nantes,
we show respectively in A,E and I the geographic representation of graph $G_{SP}$, where links correspond to the 1\% best shortest paths of the 
public transportation network. In B, F and J, we show the normalised BiCross validation errors computed for the adjacency matrix 
$X_{SP}(0 Km,5 Km)$ (grey full line) of the same graphs, for the associated random matrix $X_{SPrandom}(0 Km,5 Km)$ (dashed line). 
The selected number of structures $k_{s}$ is marked with a red rhombus. In C and D,G and H,K and L, two examples of structures 
revealed in the PT system are presented. Green dots are in-going, while red dots are out-going affiliated.}}
\label{SchemaPaths}
\end{figure*}

\begin{figure*}[htbp!]
\centering
\begin{subfigure}{.39\textwidth}
\includegraphics[width=\textwidth]{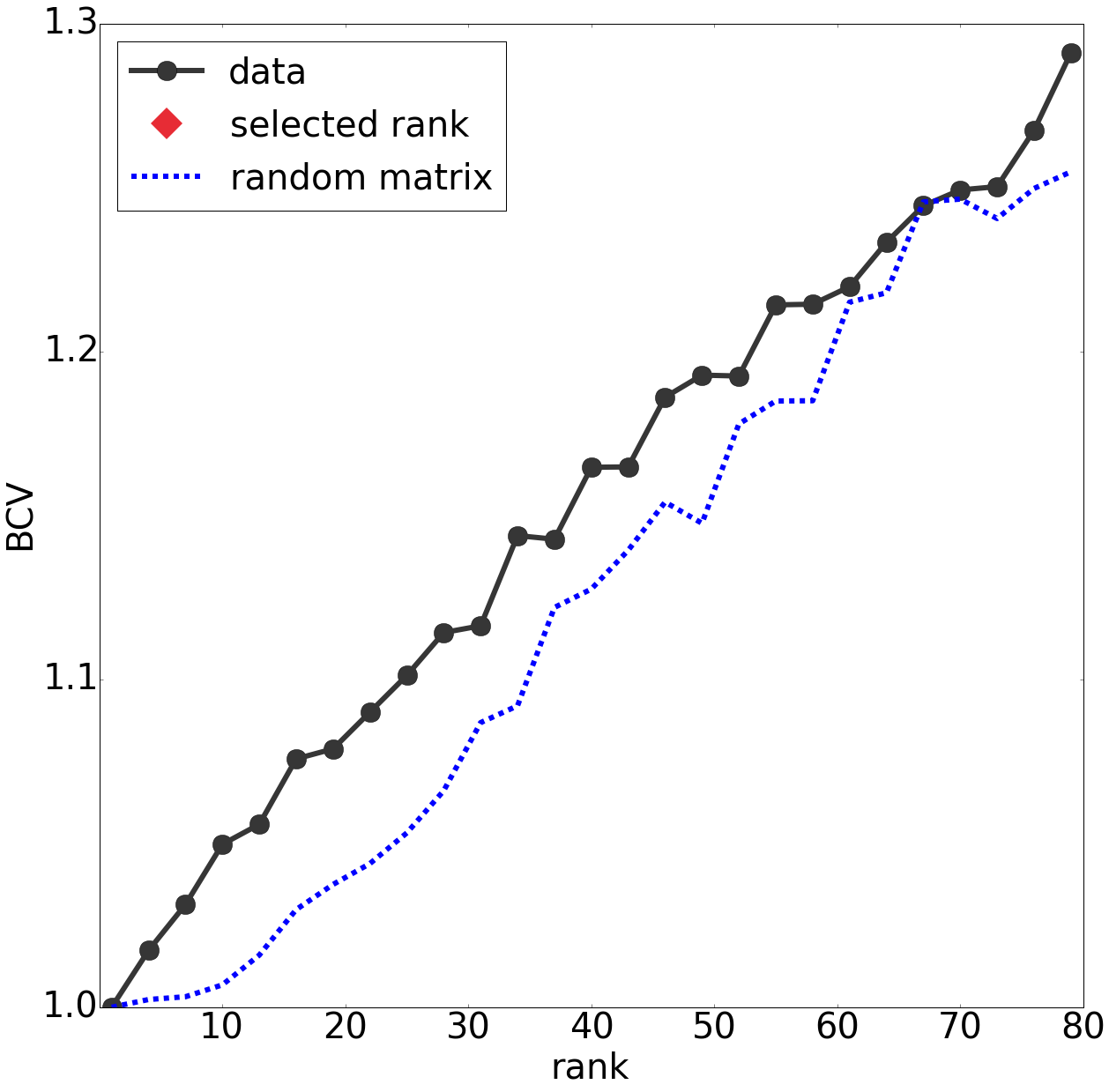}
\caption{Strasbourg,$d=5-10 Km$}
\end{subfigure}
\begin{subfigure}{.39\textwidth}
\includegraphics[width=\textwidth]{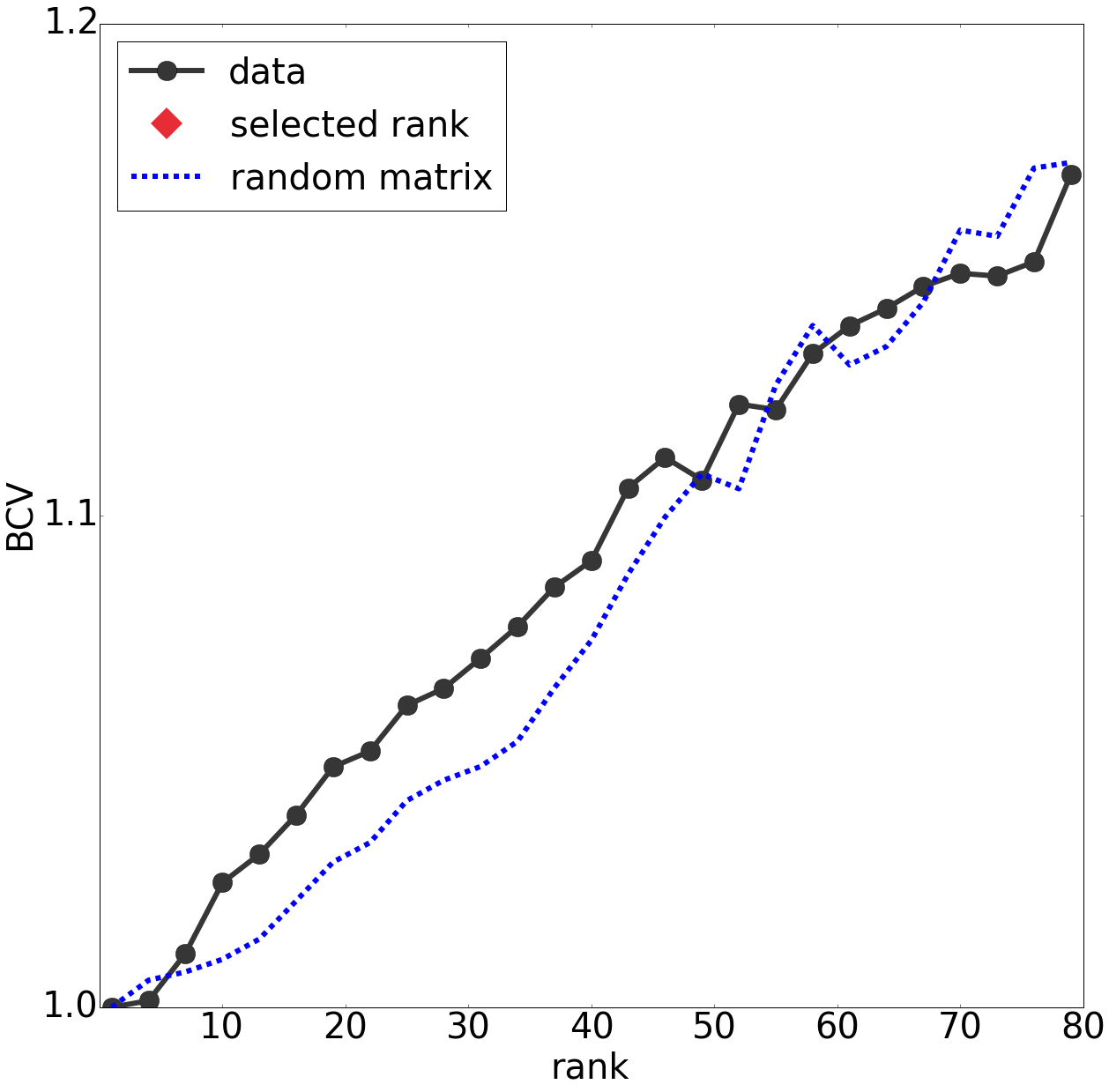}
\caption{Nantes,$d=5-10 Km$}
\end{subfigure}
\caption{For the cities of Strasbourg (a), and Nantes (b), the normalised BiCross validation error computed for the adjacency matri
$X_{SP}(5 Km,10 Km)$ (grey full line) is similar to the BiCross validation error of the associated random matrix $X_{SPrandom}(0 Km,5 Km)$ 
(dashed line). Therefore, no rank is selected and structures were not extracted.}
\label{noStructure}
\end{figure*}

\section{Comparison of the patterns detected and the commuter flows}
\begin{figure*}[htbp!]
\centering
\includegraphics[width=.8\textwidth]{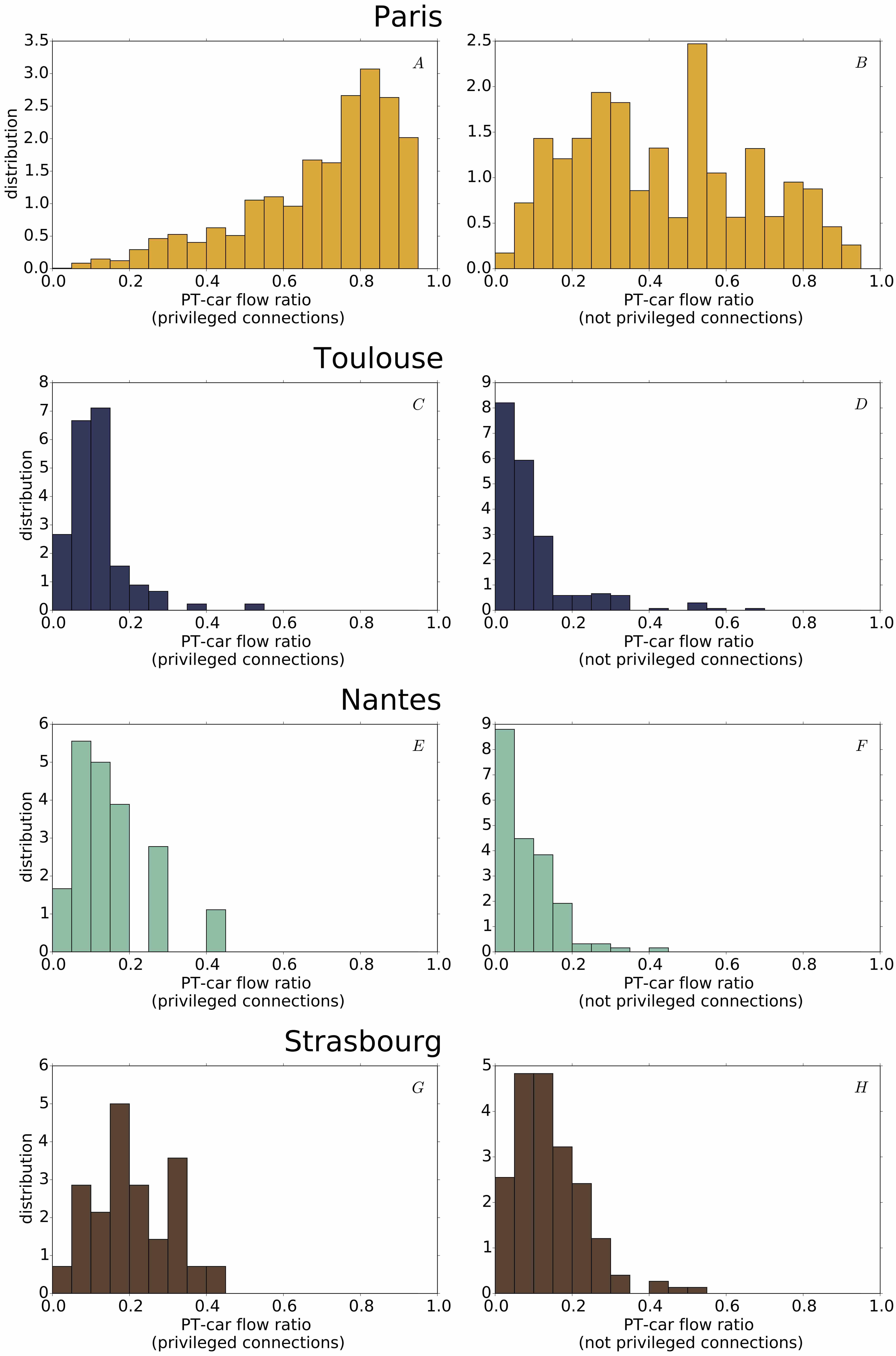}
\caption{For each city, we show the distribution of the PT-car flow ratio $f(u,v)$ when $u$ and $v$ are well (as defined in the main text)
connected (A) or badly 
(the complementary connections)  connected (B)}.
\label{ratio}
\end{figure*}

We further investigate commuters behaviour, by identifying each pair of municipalities such that a flow of commuters exists between them 
and computing the corresponding PT-car flow ratio as the fraction of commuters using PT over the total people commuting between the two cities. We then compare the cases where the two municipalities are well (Figure S\ref{ratio}, A,C,E,G) or badly (Figure S\ref{ratio}, B,D,F,H) connected by PT according to our definition, considering the distribution of the PT-car flow ratio. \\
More precisely, we consider the PT structural pattern network $G_C=(V_C,E_C)$, and the commuter flow network  $G_{com}^{TM}=(V_{com}^M,E_{com}^{TM}, W_{com}^M)$, where $M=car$ or $M=PT$; first for each edge $(u,v) \in E_C$, we compute the fraction of commuters using PT, $ f(u,v)=(W_{com}^{PT}(u,v)+W_{com}^{PT}(v,u))/ (W_{com}^{PT}(u,v)+W_{com}^{PT}(v,u)+W_{com}^{car}(v,u)+W_{com}^{car}(v,u))$.Then, we compute the same quantity for all edges $(u,v) \in E_{com}$ that are not in $E_C$. For each city, we finally look at the distribution of $f(u,v)$ for both well and badly connected municipalities (Figure S\ref{ratio}).\\

In the case of Paris agglomeration, there is a significant difference between the case of privileged connections, where the distribution is 
left-side skewed (figure S\ref{ratio} A),  and not privileged connections, where the distribution is more symmetrical (figure S\ref{ratio} A). 
This indicates that when the PT provides with good transportation according to our method, commuters prefer PT with respect to car. On the other
hand, for Toulouse, Nantes and Strasbourg agglomerations, there is significantly less difference in the distribution of the PT-car flow ratios 
for well and badly connected pairs of cities. On the one hand, this may suggest commuters tend to use the car even where good connections are
provided. On the other hand, we have to consider both that our selection was less strict for these cities, and that self loops (inter-city
connections) may play an important contribution which could not be considered here due the resolution limit of the commuter dataset.\\

\newpage

\bibliographystyle{rsos}

\end{document}